\documentclass{svjour3}                     
\smartqed  
\usepackage{graphicx}
\usepackage{amsmath,amssymb,bm}
\begin{document}

\title{Effective $\bar{K}N$ interaction with strong $\pi\Sigma$ 
dynamics constrained by chiral SU(3) symmetry
}


\author{Tetsuo Hyodo         \and
        Wolfram Weise 
}


\institute{T. Hyodo \at
              Physik-Department, Technische Universit\"at M\"unchen, 
D-85747 Garching, Germany; \\
              Yukawa Institute for Theoretical Physics,
Kyoto University, Kyoto 606--8502, Japan\\
              \email{thyodo@ph.tum.de}           
           \and
           W. Weise \at
           Physik-Department, Technische Universit\"at M\"unchen, 
D-85747 Garching, Germany
}

\date{Received: date / Accepted: date}

\maketitle

\begin{abstract}
    We derive a single-channel effective $\bar{K}N$ interaction from chiral 
    SU(3) coupled-channel dynamics, emphasizing the important role of the 
    $\pi\Sigma$ channel and the structure of the $\Lambda(1405)$ resonance. 
    The chiral low energy theorem requires strongly attractive interaction
    not only in the $\bar{K}N$ channel but also in the $\pi\Sigma$ channel. 
    As a consequence of the strong $\pi\Sigma$ dynamics, the equivalent 
    potential in single $\bar{K}N$ channel turns out to be less attractive 
    than the one used in a purely phenomenological approach.
\end{abstract}

\section{Introduction}\label{sec:intro}

The study of $\bar{K}$-nuclear systems (kaonic nuclei) attracts considerable
interest in nuclear and hadron physics. It was suggested in a 
phenomenological approach~\cite{Akaishi:2002bg} that the strongly attractive 
$\bar{K}N$ interaction could form $\bar{K}$-nuclear bound systems with 
interesting properties. Concerning the simplest $K^-pp$ system, an experiment
has performed to search for the possible bound state~\cite{Agnello:2005qj} 
while the interpretation of the results is still 
controversial~\cite{Magas:2006fn}. In recent experimental 
studies~\cite{Suzuki:2007pd,Yamazaki}, a broad peak structure in the 
$\Lambda N$ invariant mass spectrum is reported.

For a quantitative estimation of the properties of the antikaon binding in
few-nucleon systems, we need to perform rigorous few-body calculation using 
realistic potentials. The $K^-pp$ system has recently been studied in two 
approaches: Faddeev calculations in 
Refs.~\cite{Shevchenko:2006xy,Ikeda:2007nz} and a variational calculation 
with local potentials in Ref.~\cite{Yamazaki:2007cs}. For a variational 
calculation of the few-body kaonic nuclei, it is desired to construct a 
realistic $\bar{K}N$ interaction which incorporates the dynamics of the 
$\pi\Sigma$ channel. At this point, it should be noted that the relevant 
energy region for the study of kaonic nuclei is \textit{far below} the 
$\bar{K}N$ threshold, so the result would be sensitive to the extrapolation 
of the $\bar{K}N$ interaction into this energy region. The only available 
information is the invariant mass spectrum of $\pi\Sigma$ channel, which is 
dominated by the $\Lambda(1405)$ resonance. These facts indicate that we need
a careful assessment of the single-channel $\bar{K}N$ interaction together 
with a proper treatment of the $\Lambda(1405)$ resonance.

For this purpose, we rely upon the following guiding principles for the 
description of the $\bar{K}N$ scattering: chiral symmetry and coupled-channel
dynamics. Chiral symmetry of QCD determines the low energy interaction 
between the pseudoscalar meson (the Nambu-Goldstone boson) and any target 
hadron~\cite{Weinberg:1966kf}, and the importance of the coupled-channel
dynamics has been emphasized in the phenomenological study of $\bar{K}N$ 
scattering~\cite{Dalitz:1967fp}. Theoretical framework based on these 
principles has been developed as the chiral coupled-channel approach, 
reproducing successfully the $\bar{K}N$ scattering data and the properties of
the $\Lambda(1405)$ resonance~\cite{Kaiser:1995eg}. Here we derive an
effective $\bar{K}N$ interaction based on chiral 
dynamics~\cite{Hyodo:2007jq}, and discuss its phenomenological consequence in
the study of $K^-pp$ system~\cite{Dote:2008in}.

\section{Chiral low energy interaction 
and coupled-channel approach}\label{sec:interaction}

In the leading order of chiral perturbation theory, meson-baryon $s$-wave 
interaction at total energy $\sqrt{s}$ from channel $j$ to $i$ reads
\begin{align}
    V_{ij}(\sqrt{s})
    =&-\frac{C_{ij}}{4f^{2}}
    (2\sqrt{s}-M_i-M_j)\sqrt{\frac{E_i+M_i}{2M_i}}
    \sqrt{\frac{E_j+M_j}{2M_j}} 
    \sim  -\frac{C_{ij}}{4f^{2}}(\omega_i+\omega_j)
    \label{eq:WTint} ,
\end{align}
where $f$ is the pseudoscalar meson decay constant, $M_i$, $E_i$, and 
$\omega_i$ are the mass of the baryon, the energy of the baryon, and the 
energy of the meson in channel $i$, respectively. The coupling strengths
$C_{ij}$ are collected in the matrix
\begin{equation}
    C_{ij}^{I=0}
    =\begin{pmatrix}
       3 & -\sqrt{\frac{3}{2}} & \frac{3}{\sqrt{2}} & 0 \\
         & 4 & 0 & \sqrt{\frac{3}{2}} \\
	 &   & 0 & -\frac{3}{\sqrt{2}} \\
	 &   &   & 3
    \end{pmatrix}
    \nonumber ,
\end{equation}
for the $S=-1$ and $I=0$ channels in the following order : $\bar{K}N$, 
$\pi\Sigma$, $\eta\Lambda$, and $K\Xi$. The important point is that the 
properties of the interaction---sign, strength, and energy dependence---are 
strictly governed by the chiral theorem. One observes that the interactions 
in \textit{both} $\bar{K}N$ and $\pi\Sigma$ channels are attractive, which is
inevitable as far as we respect chiral symmetry. As we will see below, these 
attractive forces are so strong that pole singularities of the amplitude are 
generated for both channels. 

Since the system is strongly interacting, we need to perform nonperturbative
resummation. In Refs.~\cite{Kaiser:1995eg,Hyodo:2007jq} this has been
achieved by solving the Bethe-Salpeter equation
\begin{align}
    T_{ij}(\sqrt{s})
    =& 
    V_{ij}(\sqrt{s})+V_{il}(\sqrt{s})\,G_{l}(\sqrt{s})\,T_{lj}(\sqrt{s}),
    \label{eq:full}
\end{align}
with the interaction kernel $V_{ij}$ in Eq.~\eqref{eq:WTint} and the 
meson-baryon loop integral $G_i$ in dimensional regularization. The solution 
of Eq.~\eqref{eq:full} is given in matrix form by $T= [V^{-1}-G]^{-1}$ under 
the on-shell factorization. This form of the amplitude is also obtained in 
the N/D method by neglecting the contributions from the left-hand cut. This 
solution guarantees the unitarity of the scattering amplitude. 

\section{Structure of the $\Lambda(1405)$ resonance}\label{sec:L1405}

It has been shown that the amplitude in this framework reproduces 
the experimental observables of $\bar{K}N$ scattering very well, by properly 
choosing the subtraction constants in the loop 
function~\cite{Kaiser:1995eg,Hyodo:2002pk}. The dynamically generated 
resonance is then expressed by a pole of 
the scattering amplitude in the complex energy plane. An interesting 
observation is that the $\Lambda(1405)$ resonance is associated by two 
poles~\cite{Jido:2003cb}. Using the model given in Refs.~\cite{Hyodo:2002pk},
we find the poles at
\begin{align*}
    z_1 &= 1428 - 17 i \text{ MeV}, \quad
    z_2 = 1400 - 76 i \text{ MeV} ,
\end{align*}
which appear above $\pi\Sigma$ threshold and below $\bar{K}N$ threshold. 
Since the two poles locate close to each other, the observed spectrum 
exhibits only one bump structure, which was interpreted as a single 
resonance, the $\Lambda(1405)$. The coupling strengths of the poles to the 
$\pi\Sigma$ and $\bar{K}N$ channels are different from one to the other. 
Therefore, these poles contribute to the $\bar{K}N$ and $\pi\Sigma$ 
amplitudes with different weights, leading to the different spectral shapes 
of two amplitudes~\cite{Jido:2003cb}. 

\begin{figure}[tbp]
    \centering
    \includegraphics[width=0.5\textwidth,clip]{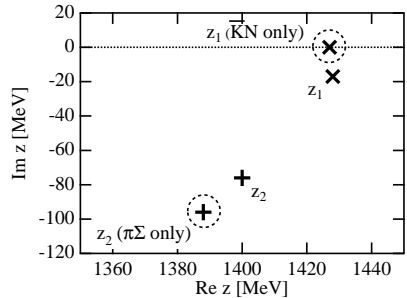}
    \caption{\label{fig:pole}
    Pole positions of the results of single-channel resummation 
    [$z_1(\bar{K}N$ only) and $z_2(\pi\Sigma$ only)] together with the poles
    in the $\bar{K}N(I=0)$ full coupled-channel amplitude ($z_1$ and $z_2$).}
\end{figure}

Here we study the origin of this 
interesting structure. In order to isolate the contribution of each channel, 
we perform the resummation of the single-channel interaction by switching off
the couplings to the other channels. This single channel $\bar{K}N$ 
interaction generates a relatively weak bound state below threshold, while 
the $\pi\Sigma$ amplitude exhibits a broad resonance above threshold:
\begin{equation}
    z_1(\bar{K}N \text{ only}) = 1427  \text{ MeV} ,
    \quad z_2(\pi\Sigma \text{ only}) = 1388 - 96 i \text{ MeV} .
    \nonumber
\end{equation}
In this way, the attractive forces in diagonal $\bar{K}N$ and $\pi\Sigma$ 
channels already generate two poles between thresholds.
We plot the positions of these poles in Fig.~\ref{fig:pole}, together with 
the poles in the full amplitude in the coupled-channel framework. The figure 
obviously suggests that the pole $z_1(\bar{K}N \text{ only})$ is the origin 
of the pole $z_1$, whereas $z_2(\pi\Sigma \text{ only})$ evolves to the pole 
$z_2$. This observation agrees with the qualitative behavior discussed in 
Ref.~\cite{Jido:2003cb}; the pole $z_1$ strongly couples to the $\bar{K}N$
channel and the pole $z_2$ to the $\pi\Sigma$ channel.

It is interesting to note that the higher energy $\bar{K}N$ channel has 
stronger attraction ($\sim 3\omega_K$) to generate a bound state, and the 
lower energy $\pi\Sigma$ channel shows the relatively weaker attraction 
($\sim 4\omega_{\pi}$), which is nevertheless strong enough to create a 
resonance. The appearance of the two poles in this energy region is caused by
the valance of the two attractive forces. As we emphasized in the previous 
section, the meson-baryon interaction is governed by the chiral low energy 
theorem. Hence, we consider that the two-pole structure of the 
$\Lambda(1405)$ is a natural consequence of chiral symmetry.

\section{Effective single-channel interaction}\label{sec:single}

Keeping the structure of the $\Lambda(1405)$ in mind, we construct an 
effective single-channel $\bar{K}N$ interaction which incorporates the 
dynamics of the other channels 2-4 ($\pi\Sigma$, $\eta N$, and $K\Xi$). We 
would like to obtain the solution $T_{11}$ of Eq.~\eqref{eq:full} by solving 
a single-channel equation with kernel interaction $V^{\text{eff}}$, namely,
\begin{align}
    T^{\text{eff}}
    =& \,V^{\text{eff}}+V^{\text{eff}}\,G_1\,T^{\text{eff}}
    = \,T_{11} .
    \nonumber 
\end{align}
Consistency with Eq.~\eqref{eq:full} requires that $V^{\text{eff}}$ be the 
sum of the bare interaction in channel 1 and the contribution 
$\tilde{V}_{11}$ from other channels:
\begin{align}
    V^{\text{eff}}
    =&\,V_{11} + \tilde{V}_{11}
    ,\quad 
    \tilde{V}_{11} 
    =\sum_{m=2}^{4}V_{1m}\,G_m\,V_{m1}
    +\sum_{m,l=2}^{4}V_{1m}\,G_m\,
    T^{(3)}_{ml}\, G_l\,
    V_{l1}
    \label{eq:Veffective} , \\
    T^{(3)}_{ml}
    =&\, 
    V_{ml}+\sum_{k=2}^{4}V_{mk}\,G_k\,T^{(3)}_{kl} ,~~
    m,l=2,3,4.
    \nonumber
\end{align}
where $T^{(3)}_{ml}$ is the $3\times 3$ matrix with indices 2-4, and 
expresses the resummation of interactions other than channel 1. Note that 
$\tilde{V}_{11}$ includes iterations of one-loop terms in channels 2-4 to all
orders, stemming from the coupled-channel dynamics. This is an exact 
transformation, as far as the $\bar{K}N$ scattering amplitude is concerned.

\begin{figure*}[tbp]
    \centering
    \includegraphics[width=0.75\textwidth,clip]{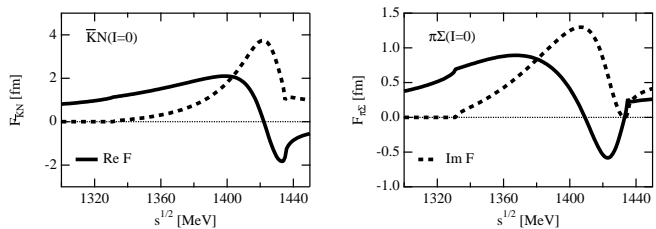}
    \caption{\label{fig:Full}
    Forward scattering amplitudes $F_{\bar{K}N}$ (left) and $F_{\pi\Sigma}$ 
    (right). Real parts are shown as solid lines and imaginary parts as 
    dashed lines. The amplitudes shown are related to the $T_{ij}$ in 
    Eq.~\eqref{eq:full} by $F_{i} = -M_i T_{ii}/(4\pi \sqrt{s})$.}
\end{figure*}%

The effective $\bar{K}N$ interaction $V^{\text{eff}}$ is calculated within a 
chiral coupled-channel model~\cite{Hyodo:2002pk}. It turns out that the 
$\pi\Sigma$ and other coupled channels enhance the strength of the 
interaction at low energy, although not by a large amount. The primary effect
of the coupled channels is found in the energy dependence of the 
interaction kernel. In the left panel of Fig.~\ref{fig:Full}, we show the 
result of $\bar{K}N$ scattering amplitude $T^{\text{eff}}$, which is obtained
by solving the single-channel scattering equation with $V^{\text{eff}}$. The 
full amplitude in the $\pi\Sigma$ channel is plotted in the right panel for 
comparison. It is remarkable that the resonance structure in the $\bar{K}N$ 
channel is observed at around 1420 MeV, higher than the nominal position of 
the $\Lambda(1405)$. What is experimentally observed is the spectrum in the 
$\pi\Sigma$ channel, whose peak is in fact located close to 1405 
MeV\footnote{Strictly speaking, the observed $\pi\Sigma$ spectrum should be 
described by the sum of the possible initial states 
$\sum_{i}C_i|T_{i\to \pi\Sigma}|^2$~\cite{Jido:2003cb}. To 
obtain the relative weights $C_{i}$, we need to introduce a model for each 
reaction process as done in Refs.~\cite{Jido:2003cb}.}. The
discrepancy of the shapes of the two amplitudes is a consequence of the 
existence of two poles with different coupling strengths~\cite{Jido:2003cb}.

The deviation of the resonance position has a large impact to the 
single-channel $\bar{K}N$ potential, which should be constructed so as 
to reproduce the scattering amplitude in $\bar{K}N$ channel. Measuring from 
the $\bar{K}N$ threshold ($\sim 1435$ MeV) we find the ``binding energy'' of 
the $\Lambda(1405)$ as $\sim 15$ MeV, not as the nominal value of $\sim 30$ 
MeV. This shift of the binding energy apparently reduces the strength of
the potential.

\section{Equivalent local potential}\label{sec:potential}

Next we construct an equivalent local $\bar{K}N$ potential in coordinate 
space. We consider an $s$-wave antikaon-nucleon system in nonrelativistic 
quantum mechanics, whose radial wave function $u(r)$ follows the 
Schr\"odinger equation with the potential $U(r,E)$. As explained in detail in
Ref.~\cite{Hyodo:2007jq}, the local potential $U(r,E)$ has been constructed
such that the scattering amplitude in coupled-channel approach is reproduced 
in this system. Note that this is not an exact transformation, since it is 
not guaranteed that a simple local potential can reproduce the complicated 
coupled-channel dynamics. Nevertheless, we have constructed a complex and 
energy-dependent $\bar{K}N$ potential with the gaussian form of the spatial 
distribution, which well reproduces the coupled-channel results.

It is instructive to compare our potential with the phenomenological 
Akaishi-Yamazaki (AY) potential~\cite{Akaishi:2002bg,Yamazaki:2007cs}, whose 
form in $\bar{K}N$ and $\pi\Sigma$ channels is given by
\begin{equation}
    v_{ij}(r)
    =\begin{pmatrix}
       -436 & -412 \\
         -412 & 0 & 
    \end{pmatrix}
    \exp [-(r/b)^2] \quad \text{[MeV]}
    \nonumber ,
\end{equation}
with $b\sim 0.66$ fm. The interaction is qualitatively different from the 
chiral interaction, namely, the absence of a direct $\pi\Sigma\to \pi\Sigma$ 
coupling. The strength of the phenomenological interaction is determined by 
assuming the resonance position in $\bar{K}N$ channel is the same with the 
PDG value of the $\Lambda(1405)$ resonance. In this framework, the 
$\Lambda(1405)$ is described as a Feshbach resonance: $\bar{K}N$ quasibound 
state embedded in the $\pi\Sigma$ continuum.

Chiral SU(3) dynamics, on the other hand, leads to the quasibound structure 
in the $\bar{K}N$ system at around 1420 MeV, because of the strong 
$\pi\Sigma$ diagonal interaction. As a result, the equivalent local
$\bar{K}N$ potential is less attractive. This is inevitable, as the chiral 
low energy theorem restricts the structure of the meson-baryon interaction. 
In this framework, as in the AY potential, the driving force to generate the 
$\Lambda(1405)$ is the $\bar{K}N$ attraction, but the chiral low energy 
theorem requires the $\pi\Sigma$ continuum to be also strongly interacting. 
As a consequence, we observe an interesting structure: $\bar{K}N$ quasibound 
state embedded in the \textit{resonating} $\pi\Sigma$ continuum.

It should be however noted that both chiral and phenomenological amplitudes 
behave similarly \textit{above} the $\bar{K}N$ threshold, since the 
potentials are adjusted to describe experimental data. The difference appears
in the treatment of the $\pi\Sigma$ interaction and extrapolation of the 
$\bar{K}N$ amplitude to the subthreshold energy region. In other words, the 
existing experimental database, by itself, is not sufficient to constrain the
$\bar{K}N$ interaction in the energy region relevant for the antikaon-nucleon
physics.

\section{Summary}\label{sec:summary}

We have derived an effective $\bar{K}N$ interaction based on chiral low 
energy theorem and the coupled-channel dynamics. We show that the 
model-independent chiral interaction leads to the strongly interacting 
$\pi\Sigma$-$\bar{K}N$ system, in which the $\Lambda(1405)$ is described as 
the $\bar{K}N$ quasibound state embedded in the \textit{resonating} 
$\pi\Sigma$ continuum. We construct an equivalent local potential in single 
$\bar{K}N$ channel, which represents the effect of coupled-channel dynamics
through the imaginary part and energy dependence. As a consequence of the 
strong $\pi\Sigma$ dynamics, the resulting potential is less attractive than 
the purely phenomenological potential in the energy range relevant to the 
discussion of deeply bound kaonic nuclei.

It is worth emphasizing that there is no direct experimental constraint on 
the $\bar{K}N$ amplitude below threshold. We have to extrapolate $\bar{K}N$ 
interaction calibrated by scattering data above threshold, down to the 
relevant energy scale. Here we utilize the principle of chiral SU(3) symmetry
in order to reduce the ambiguity of the extrapolation. The precise knowledge 
of the threshold $\bar{K}N$ data and the $\pi\Sigma$ mass spectrum will be 
important for the prediction of the kaonic nuclei.

\begin{acknowledgements}
    This project is partially supported by BMBF, GSI, by the DFG excellence 
    cluster ``Origin and Structure of the Universe.", by the Japan Society 
    for the Promotion of Science (JSPS), and by the Grant for Scientific 
    Research (No.\ 19853500) from the Ministry of Education, Culture, Sports,
    Science and Technology (MEXT) of Japan. This research is part of the 
    Yukawa International Program for Quark-Hadron Science. 
\end{acknowledgements}


\begin{thebibliography}{}
%
%
    
\bibitem{Akaishi:2002bg}
Y.~Akaishi and T.~Yamazaki,
\newblock Phys. Rev. C {\bf 65}, 044005 (2002).

\bibitem{Agnello:2005qj}
M.~Agnello {\em et~al.}, (FINUDA collaboration),
\newblock Phys. Rev. Lett. {\bf 94}, 212303 (2005).

\bibitem{Magas:2006fn}
V.~K. Magas, E.~Oset, A.~Ramos, and H.~Toki,
\newblock Phys. Rev. C {\bf 74}, 025206 (2006).

\bibitem{Suzuki:2007pd}
  T.~Suzuki {\it et al.}  [KEK-PS E549 Collaboration],
  arXiv:0711.4943 [nucl-ex].

\bibitem{Yamazaki}
  T.~Yamazaki {\it et al.}, arXiv:0810.5182 [nucl-ex], in this proceedings.
  
\bibitem{Shevchenko:2006xy}
N.~V. Shevchenko, A.~Gal, and J.~Mares,
\newblock Phys. Rev. Lett. {\bf 98}, 082301 (2007);
N.~V. Shevchenko, A.~Gal, J.~Mares, and J.~Revai,
Phys. Rev. C {\bf 76}, 044004 (2007).

\bibitem{Ikeda:2007nz}
Y.~Ikeda and T.~Sato,
\newblock Phys. Rev. C {\bf 76}, 035203 (2007).

\bibitem{Yamazaki:2007cs}
T.~Yamazaki and Y.~Akaishi,
\newblock Phys. Rev. C {\bf 76}, 045201 (2007).

\bibitem{Weinberg:1966kf}
S.~Weinberg,
\newblock Phys. Rev. Lett. {\bf 17}, 616 (1966).
Y.~Tomozawa,
\newblock Nuovo Cim. {\bf 46A}, 707 (1966).

\bibitem{Dalitz:1967fp}
  R.~H.~Dalitz, T.~C.~Wong and G.~Rajasekaran,
  Phys.\ Rev.\  {\bf 153}, 1617 (1967).
  
\bibitem{Kaiser:1995eg}
N.~Kaiser, P.~B. Siegel, and W.~Weise,
\newblock Nucl. Phys. {\bf A594}, 325 (1995);
E.~Oset and A.~Ramos,
\newblock Nucl. Phys. {\bf A635}, 99 (1998);
M.~F.~M. Lutz and E.~E. Kolomeitsev,
\newblock Nucl. Phys. {\bf A700}, 193 (2002);
J.~A. Oller and U.~G. Meissner,
\newblock Phys. Lett. {\bf B500}, 263 (2001).

\bibitem{Hyodo:2007jq}
 T.~Hyodo and W.~Weise,
 \newblock Phys. Rev. C {\bf 77}, 035204 (2008). 

\bibitem{Dote:2008in}
A.~Dote, T.~Hyodo, and W.~Weise,
\newblock Nucl. Phys. {\bf A804}, 197 (2008); 
arXiv:0806.4917.

\bibitem{Hyodo:2002pk}
T.~Hyodo, S.~I. Nam, D.~Jido, and A.~Hosaka,
\newblock Phys. Rev. C {\bf 68}, 018201 (2003);
Prog. Theor. Phys. {\bf 112}, 73 (2004).

\bibitem{Jido:2003cb}
D.~Jido, J.~A. Oller, E.~Oset, A.~Ramos, and U.~G. Meissner,
\newblock Nucl. Phys. {\bf A725}, 181 (2003);
T.~Hyodo, A.~Hosaka, E.~Oset, A.~Ramos, and M.~J. V.~Vacas,
\newblock Phys. Rev. C {\bf 68}, 065203 (2003);
V.~K. Magas, E.~Oset, and A.~Ramos,
\newblock Phys. Rev. Lett. {\bf 95}, 052301 (2005).
   
\end{thebibliography}


\end{document}